\documentclass[aps,twocolumn,floats,prl,nofootinbib,superscriptaddress,10pt,longbibliography]{revtex4-1}
\usepackage[utf8]{inputenc}
\usepackage{graphicx}
\usepackage{dcolumn}
\usepackage{bm}
\usepackage{hyperref}
\usepackage{lipsum}
\usepackage{xcolor}
\usepackage{calc}
\usepackage{accents}
\usepackage{comment}
\usepackage{float}
\usepackage{diagbox}
\usepackage{soul}
\usepackage{ulem}

\newcommand\mnras{{\rm MNRAS}}


\begin{document}

\title{Late Time Modification of Structure Growth and the S8 Tension}

\author{Meng-Xiang Lin}
\email{mxlin@sas.upenn.edu}
\affiliation{Center for Particle Cosmology, Department of Physics and Astronomy, University of Pennsylvania, Philadelphia, PA 19104, USA}

\author{Bhuvnesh Jain}
\affiliation{Center for Particle Cosmology, Department of Physics and Astronomy, University of Pennsylvania, Philadelphia, PA 19104, USA}

\author{Marco Raveri}
\affiliation{Department of Physics, INFN and INAF, University of Genova, Via Dodecaneso 33, 16146, Italy}

\author{Eric J. Baxter}
\affiliation{Institute for Astronomy, University of Hawai`i, 2680 Woodlawn Drive, Honolulu, HI 96822, USA}

\author{Chihway Chang}
\affiliation{Kavli Institute for Cosmological Physics, University of Chicago, Chicago, IL 60637, USA}

\author{Marco Gatti}
\affiliation{Center for Particle Cosmology, Department of Physics and Astronomy, University of Pennsylvania, Philadelphia, PA 19104, USA}

\author{Sujeong Lee}
\affiliation{Jet Propulsion Laboratory, California Institute of Technology, Pasadena, CA 91109, USA}

\author{Jessica Muir}
\affiliation{Perimeter Institute for Theoretical Physics, 31 Caroline St N, Waterloo, ON N2L 2Y5, Canada}

\begin{abstract}
    The $S_8$ tension between low-redshift galaxy surveys and the primary CMB signals a possible breakdown of the $\Lambda$CDM model. 
    Recently differing results have been obtained using low-redshift galaxy surveys and the higher redshifts probed by CMB lensing, motivating a possible time-dependent modification to the growth of structure. 
    We investigate a simple phenomenological model in which the growth of structure deviates from the $\Lambda$CDM prediction at late times, in particular as a simple function of the dark energy density. Fitting to galaxy lensing, CMB lensing, BAO, and Supernovae datasets, we find significant evidence - 2.5 - 3$\sigma$, depending on analysis choices - for a non-zero value of the parameter quantifying a deviation from $\Lambda$CDM. 
    The preferred model, which has a slower growth of structure below $z\sim 1$, improves the joint fit to the data over $\Lambda$CDM.  
    While the overall fit is improved, there is weak evidence for galaxy and CMB lensing favoring different changes in the growth of structure.   
\end{abstract}

\maketitle

\textbf{Introduction.}
In current empirical cosmology two ``cosmic tensions'' have been actively pursued: the Hubble tension and the $S_8$ tension. The Hubble tension refers to the present-day expansion rate being faster than predicted by the $\Lambda$CDM model (the standard, minimal cosmological model with a Cosmological constant $\Lambda$ plus Cold Dark Matter dominating the energy density) \cite{Planck:2018vyg, Riess:2021jrx}. The prediction is obtained by using the CMB temperature and polarization measurements at $z\approx 1100$ and extrapolating in time to the present using the expansion rate expected in the $\Lambda$CDM cosmology. 
The $S_8$ tension (aka $\sigma_8$ tension) refers to the amplitude of matter density fluctuations at the present being smaller than predicted in $\Lambda$CDM, again based on the measurements of the CMB \cite{DES:2021wwk}. The statistical significance of the $S_8$ tension is only at the $2-3\sigma$ level, and different galaxy survey measurements -- in particular weak lensing (WL) and redshift space distortions (RSD) -- have some spread in the inferred value. Nevertheless it has held up for several years, across experiments, and has major implications for cosmology and fundamental physics, so it has rightly received attention and scrutiny. 

WL measurements from three powerful galaxy surveys, Dark Energy Survey (DES), Kilo-degree Survey (KiDS), and Hyper Suprime-Cam Subaru Strategic Program (HSC SSP), have obtained consistent results in the $S_8-\Omega_m$ plane that show a $2-3\sigma$ tension with Planck \cite{DES:2021wwk,KiDS:2020suj,Heymans:2020gsg,Dalal:2023olq,Amon:2022ycy,Arico:2023ocu}.  
Recent measurements from the Atacama Cosmology Telescope (ACT) \cite{ACT:2023kun} and re-analysis of the Planck data \cite{Carron:2022eyg,Rosenberg:2022sdy} have provided an interesting viewing angle to the $S_8$ tension. These projects find that their lensing measurements are consistent with the primary CMB fluctuations and in $\sim 2\sigma$ tension with galaxy lensing.  These different findings from CMB lensing and galaxy lensing, if not just a statistical fluctuation, suggest that either set of measurements has a bias in the inferred amplitude, or that the growth history of the universe has some characteristic length or time scales \cite{Garcia-Garcia:2021unp,White:2021yvw,DES:2022ccp,Pourtsidou:2016ico,Amon:2022azi,Preston:2023uup,Poulin:2022sgp,Nguyen:2023fip,Wen:2023bcj,Esposito:2022plo,Naidoo:2022rda,Adil:2023jtu,Gangopadhyay:2023nli,Sakr:2018new,Sakr:2023bms}. 

In this {\it paper}, we explore the latter possibility. We note that the lensing kernel of galaxy surveys peaks at $z\approx 0.5$, while the CMB lensing kernel spans from recombination to today and has a very wide peak at $z\sim 1-3$.
Coincidentally, the epoch of the dark energy-matter equality occurs at $z\sim0.7$ --- assuming a $\Lambda$CDM expansion history and Supernovae observations~\cite{Brout:2022vxf} --- which lies between the two sensitivity peaks. 
Exploiting this possible coincidence, we introduce a two-parameter model in which deviations of growth from $\Lambda$CDM track the evolution of the dark energy density, and the behavior of the background is fixed to the $\Lambda$CDM one.  We use this model to study a possible resolution to the apparent disagreement between the galaxy lensing and CMB lensing measurements. For simplicity, we do not consider other late-time probes of growth of structure, including galaxy clustering in redshift or angular space and galaxy clusters which are broadly consistent with lensing measurements but have different sources of uncertainty. 

\begin{figure}[]
    \centering
    \includegraphics[width=0.99\columnwidth]{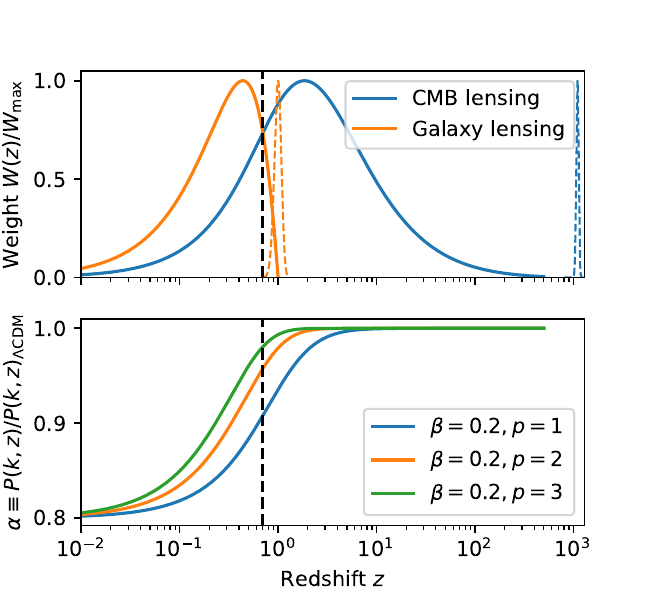}
    \caption{
    {\it Top panel}: the redshift distributions of CMB and galaxy lensing kernels (solid lines) and the sources (dashed lines).
    {\it Bottom panel}: the parameterization of DETG model Eq.~(\ref{eq:alpha}). 
    For illustration purposes, here we choose $\Omega_{\rm DE}^0=0.7$.
    The vertical dashed lines indicate $z=0.7$ where dark energy density starts to dominate.
    }
    \label{fig:alpha}
\end{figure}

\textbf{The Model.}
We introduce a phenomenological model that rescales the structure growth as a function of redshift. We define the linear matter power spectrum with the modified growth as follows:
\begin{equation}\label{eq:alpha}
    \alpha(z) \equiv \frac{P(k,z)}{P(k,z)_{\rm \Lambda CDM}}= 1-\beta\left(\frac{\Omega_{\rm DE}(z)}{\Omega_{\rm DE}^0}\right)^{p}.
\end{equation}
where $P(k,z)$ is the linear matter power spectrum, $\Omega_{\rm DE}(z)$ is the fractional energy density of dark energy at a certain time, and $\Omega_{\rm DE}^0$ is its value today. We then use \texttt{Halofit}\cite{Takahashi:2012em,Bird:2011rb} to compute the nonlinear power spectrum.
Here we assume \texttt{Halofit} can still capture the main features of the non-linear enhancement under the modification. For a more accurate analysis, one needs to do additional N-body simulations which we defer to future work.

When $\beta=0$, $\alpha = 1$ and we recover the $\Lambda$CDM model.
We illustrate this parameterization in Fig.~\ref{fig:alpha} together with the redshift distributions of CMB and galaxy lensing kernels following \cite{ACT:2023kun}.
The rescaling is tied to the time dependence of dark energy density, so it mainly changes the late-time structure growth while leaving the early-time physics unchanged. Note that the parameter $p$ alters the effective redshift at which the modification occurs, allowing us to tune the sensitivity of CMB lensing to the change. 
In the analysis below, we fix $p$ to several discrete values and mainly focus on the simplest case $p=1$.
Hereafter we refer to this model as the Dark Energy Tracking Growth (DETG) model.

In the literature, the ``growth index parameter'' $\gamma$ has been introduced to characterize gravity theories with a single parameter~\cite{Linder:2005in}. It is given by 
\begin{equation}
    g(a) = \exp\left[ \int_0^a \frac{da'}{a'} \left(\Omega_m(a')^\gamma -1\right) \right]
\end{equation}
where $\Omega_m(a)$ is the fractional energy density of matter, $g=\delta/a$ and $\delta=\delta\rho_m/\rho_m$ is the linear density perturbation of matter.
Our parameterization of eq.~(\ref{eq:alpha}) can be related to the $\gamma$ parameter above:
\begin{equation}
    \Omega_m(a)^\gamma = \Omega_m(a)^{\gamma_{\rm LCDM}} + \frac{1}{2}\frac{d\ln\alpha(a)}{d\ln a} \ .
\end{equation}
Our DETG model is somewhat similar in implementation to other phenomenological parameterizations of non-standard growth, such as those studied in e.g. Refs~\cite{Nguyen:2023fip,Wen:2023bcj,Brieden:2022lsd,DES:2022ccp}, but represents a more targeted search in that it imposes a redshift dependence 
associated with the evolution of dark energy density
and introduces only one free parameter when fixing $p=1$.

\textbf{Datasets.}
In order to assess the ability of the DETG model to reconcile the different $S_8$ constraints between galaxy lensing and CMB lensing, we use both sets of measurements as well as their cross-correlations. We use the DES Year 3 dataset for galaxy lensing, which bins the source galaxies into 4 redshift bins with the bulk of the sample over the range $0.2 \lesssim z \lesssim 1.3$ which are lensed by the foreground mass. The cross-correlation with CMB lensing is determined by the overlap of the galaxy lensing and CMB lensing redshift kernels. For additional constraints on the distance-redshift relation, we use Supernovae and BAO measurements. In particular, we employ the following datasets in this work:
\begin{itemize}
    \item DES\&xco: DES Year 3 shear$\times$shear correlations (for DES galaxies) and shear$\times$CMB lensing 2pt correlations\cite{DES:2021bvc,DES:2021vln,DES:2022xxr}. The CMB lensing is from SPT+Planck and the cross-correlation has a lower signal-to-noise ratio. 
    \item CMBlens: Planck 2018 \cite{Planck:2018lbu} and ACT DR6 \cite{ACT:2023kun,ACT:2023dou} lensing potential power spectra.
    \item BAO: Distance measurements from 6DFGS at $z$=0.106 \cite{Beutler:2011hx}, SDSS DR7 main galaxy sample at $z$=0.15 \cite{Ross:2014qpa}, and BOSS DR 12 \& 16 at $z$=0.38, 0.51, 0.68 \cite{BOSS:2016wmc,Bautista:2020ahg,Gil-Marin:2020bct}.
    \item SN: Pantheon+ supernovae dataset of relative luminoisty distances \cite{Brout:2022vxf}.
    \item All = DES\&xco + CMBlens + BAO + SN.
\end{itemize} 
For additional tests of the robustness of the results, we also add the primary CMB likelihood
\begin{itemize}
    \item CMB primary: high-$\ell$ Planck 2018 \texttt{[Plik]} temperature and polarization (TT+TE+EE) and low-E power spectra \cite{Planck:2019nip}.
\end{itemize}
We do not attempt to model the modifications of primary CMB anisotropies due to lensing and ISW effects of our modified late-time growth, instead, we marginalize over the $A_{\rm lens}$ parameter to accommodate the modified lensing effect and leave out the temperature data at low-$\ell$ that may be impacted by the late time ISW effect. The optical depth $\tau$ is also varied in the primary CMB analysis.

\textbf{Results.}
We perform Markov Chain Monte Carlo (MCMC) analyses with the \texttt{Polychord} algorithm~\cite{Handley:2015a,Handley:2015b} using \texttt{COSMOSiS}\cite{Zuntz:2014csq}\footnote{https://cosmosis.readthedocs.io/en/latest/index.html}. 
We use \texttt{CAMB}\cite{Lewis:1999bs} for cosmological calculations and \texttt{GetDist}\cite{Lewis:2019xzd} for MCMC chain analyses.
In the analyses, we fix $p$ to several discrete constant values. 
In addition to the standard $\Lambda$CDM parameters, we have one more varying parameter $\beta$ for which we impose a flat prior $-1.0<\beta<1.0$. 
In our baseline analysis, we have flat priors for $\Lambda$CDM parameters: $0.1<\Omega_m<0.9$, $55<H_0<91$, $0.03<\Omega_b<0.07$, $0.9<n_s<1.0$, $0.3<\sigma_8<2.0$.
For the robustness test, we also try a Gaussian prior on $n_s\sim \mathcal{N}(0.96,0.02)$ for the case the primary CMB likelihood is not included.

\begin{figure}[!ht]
    \centering
    \includegraphics[width=0.99\columnwidth]{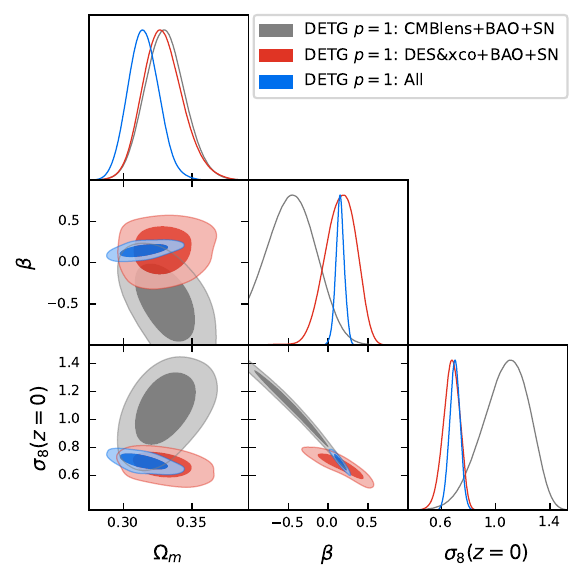}
    \caption{
    The marginalized joint posterior of parameters of the DETG $p=1$ model for different datasets. 
    Here $\sigma_8(z=0)=\sigma_8^{\rm \Lambda CDM}(1-\beta)$ is inferred from the rescaled matter power spectrum $P(k,z=0)$.
    The darker and lighter shades correspond respectively to the 68\% C.L. and 95\% C.L. 
    }
    \label{fig:p=1}
\end{figure}

\begin{figure}[!ht]
    \centering
    \includegraphics[width=0.99\columnwidth]{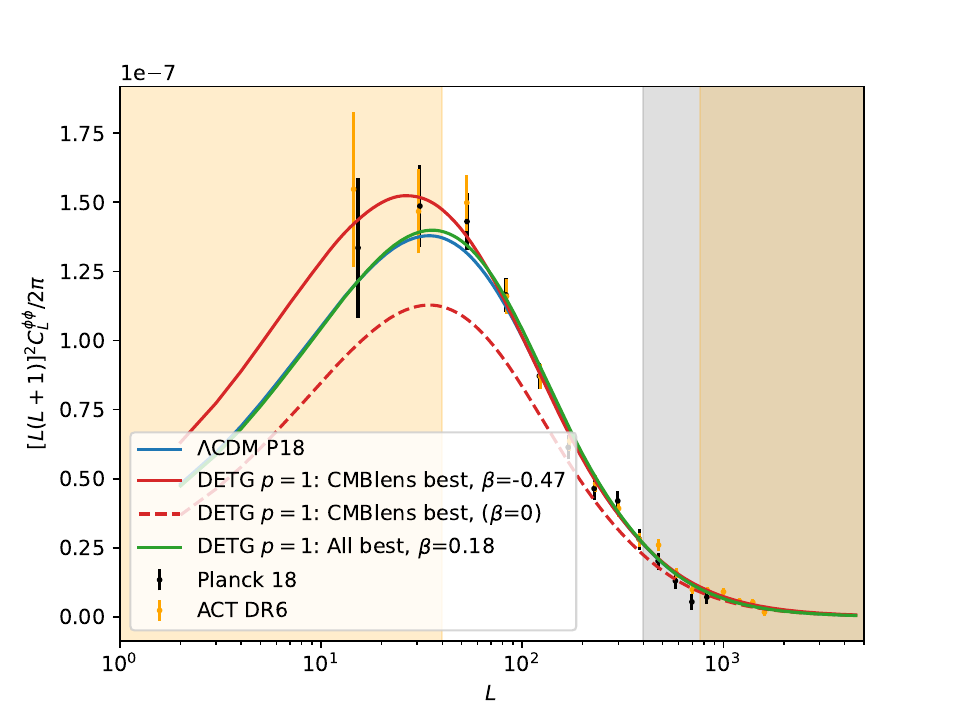}
    \caption{
    Lensing power spectra of different models along with Planck 2018 and ACT DR6 data. For DETG $p=1$ model, the solid lines are maximum likelihood models with different datasets, while the dashed red line has all parameters the same as the solid red line but with $\beta=0$. The maximum likelihood $\Lambda$CDM model given by Planck 2018 \cite{Planck:2018vyg} (the solid blue) is also shown for comparison. 
    The gray and orange shades indicate the excluded data ranges for Planck ($L>400$) and ACT ($L<40$, $L>763$) respectively.
    }
    \label{fig:phiphi}
\end{figure}

\begin{table}[!ht]
    \centering
    \begin{tabular}{c|c|c}
        \hline
        model               & $\Lambda$CDM & DETG $p=1$  \\
        \hline
        $\Omega_m$          & 0.300 (0.301$\pm$0.009) & 0.321 (0.315$\pm$0.011)  \\
        $H_0$               & 72.64 (68.68$\pm$6.59) & 66.85 (66.98$\pm$6.13) \\
        $\Omega_b$          & 0.0528 (0.0484$\pm$0.0093) & 0.0464 (0.0452$\pm$0.0090)  \\
        $n_s$               & 0.907 (0.937$\pm$0.027) & 0.909 (0.932$\pm$0.023)  \\
        $\sigma_8(z=0)$     & 0.788 (0.792$\pm$0.015) & 0.673 (0.699$\pm$0.038) \\
        $\beta$             & 0 & 0.182 (0.149$\pm$0.052) \\
        \hline
        $\Delta\chi^2_{\rm DES\&xco}$  & 0 & -2.9  \\
        $\Delta\chi^2_{\rm CMBlens}$ & 0 & 0.2  \\
        $\Delta\chi^2_{\rm BAO}$     & 0 & -1.1  \\
        $\Delta\chi^2_{\rm SN}$      & 0 & -3.3  \\
        $\Delta\chi^2_{\rm prior}$   & 0 & -0.2  \\
        \hline
        $\Delta\chi^2_{\rm tot}$     & 0 & -7.2 \\
        \hline
    \end{tabular}
    \caption{Maximum likelihood parameters and constraints (mean and the 68\% C.L. lower and upper limits) for different models fitting to All datasets. $\Delta\chi^2$ values are quoted relative to the maximum likelihood $\Lambda$CDM model.
    }
    \label{tab:best}
\end{table}

We show the results of our simplest DETG $p=1$ model in Fig.~\ref{fig:p=1}. DES\&xco data, in combination with the probes of the expansion history, favor a positive $\beta$ as we expect due to the slower late-time structure growth compared to $\Lambda$CDM.
On the other hand, CMB lensing data favor a negative $\beta$. 
Note that in the linear regime, CMB lensing by itself has a degeneracy between $\beta$ and $\sigma_8$, whereas the multiple redshift bins of DES break that degeneracy. Nonlinear evolution also helps break it for both probes. In addition, the time-dependent modification induces scale-dependence on CMB lensing which helps break the degeneracy as well.
In Fig.~\ref{fig:phiphi} we show the lensing power spectra of different maximum likelihood models in solid lines. We see that Planck and ACT lensing power spectra have relatively higher amplitudes at low-$\ell$ compared to $\Lambda$CDM. 
For the DETG $p=1$ model, varying $\beta$ has some impact on the redshift range at which the CMB lensing kernel peaks, see Fig.~\ref{eq:alpha}.
To illustrate the impact of $\beta$, we also show the dashed red line which has all the parameters the same as the maximum likelihood solid red line but with $\beta=0$. 
For the overall lensing power spectrum amplitude, $\beta$ compensates the effects of adjusting $\Lambda$CDM parameters.
Additionally, the low-$\ell$ CMB lensing power is more sensitive to low redshift structure growth, hence a negative $\beta$ enhances the low-$\ell$ CMB lensing power spectra relative to the high-$\ell$ multipoles, which is favored by the data.
However, this preference is weak as the 2$\sigma$ contour extends to the positive $\beta$ as shown in Fig.~\ref{fig:p=1}. When combining CMB lensing with DES\&xco data, it leads to a positive $\beta$ since their different $S_8$ constraints at different redshifts prefer a suppression of late-time growth. The $\Lambda$CDM value $\beta=0$ is excluded at $2.9\sigma$. 

Table \ref{tab:best} shows the parameters of the maximum likelihood models fitting to All datasets. Compared to $\Lambda$CDM, DETG $p=1$ model has a better fit by $\Delta\chi^2_{\rm tot}=-7.2$ at the expense of only one more parameter.
This improvement in $\chi^2$ is roughly compatible with the $2.9 \sigma$ deviation from $\Lambda$CDM we see in $\beta$. 
Notice that the best-fit $n_s$ is close to the lower bound of the flat prior; in fact, even after combining DES\&xco and CMB lensing data $n_s$ is not well constrained, and the preference for low $n_s$ can also be seen in the DES-Y3 cosmic shear constraints \cite{DES:2021vln}.

\begin{figure}[!ht]
    \centering
    \includegraphics[width=0.99\columnwidth]{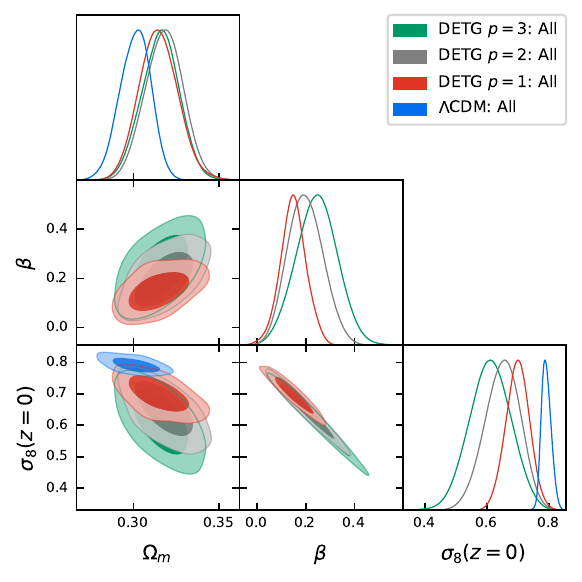}
    \caption{
    The marginalized joint posterior of parameters of different models for All datasets. The darker and lighter shades correspond respectively to the 68\% C.L. and 95\% C.L. 
    }
    \label{fig:data-all}
\end{figure}

The posteriors of DETG models with different values of the parameter $p$ fit to All datasets are shown in Fig.~\ref{fig:data-all} along with $\Lambda$CDM results.
As $p$ goes to higher values, the modification of structure growth affects lower redshifts (see Fig.~\ref{eq:alpha}), and its impact on CMB lensing becomes less important. Therefore CMBlens data have less constraining power on $\beta$, and almost no constraint for the $p=3$ model.  The posterior of $\beta$ extends to higher values for a larger $p$. 

\begin{figure}[!ht]
    \centering
    \includegraphics[width=0.99\columnwidth]{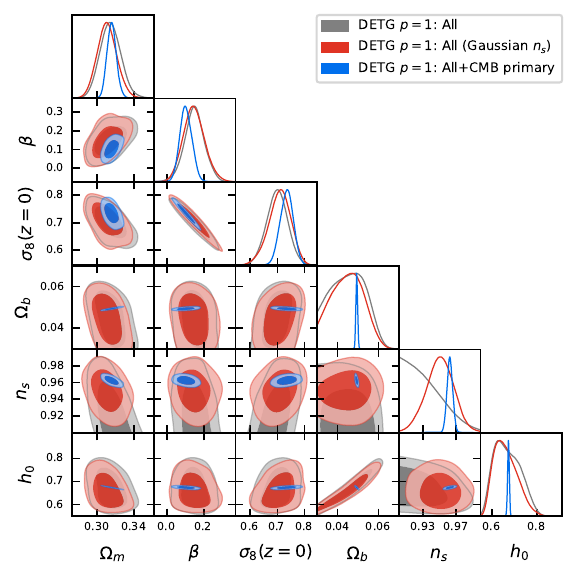}
    \caption{
    The marginalized joint posterior of parameters of the DETG $p=1$ model for different priors/datasets. 
    The darker and lighter shades correspond respectively to the 68\% C.L. and 95\% C.L. 
    }
    \label{fig:p=1_bigtri}
\end{figure}

To test the robustness of our results against primary CMB power spectra, we carry out two additional variations of our analyses with the DETG $p=1$ model: 1. using a Gaussian prior on $n_s\sim\mathcal{N}(0.96,0.02)$; 2. adding CMB primary likelihood (as noted above, with $A_{\rm lens}$ as a free parameter to account for the possible modified lensing effect of the DETG model).
The posterior distributions are shown in Fig.~\ref{fig:p=1_bigtri}.
Both are consistent with our baseline results and show evidence of deviation from $\Lambda$CDM, at $2.6\sigma$ and $3.1\sigma$ respectively.

\textbf{Conclusion and Discussion.}
This work is motivated by the difference in matter clustering inferred from CMB lensing versus galaxy lensing. While a new (or under-estimated) source of systematic uncertainty in one or both datasets is a possible explanation, here we explore the possibility that the late-time growth of structure deviates from the standard $\Lambda$CDM model. We use measurements of CMB lensing, galaxy lensing, and their cross-correlation to constrain an ad-hoc but simple model for the modification of late-time growth. Our model, which we call Dark Energy Tracking Growth (DETG), modifies the growth rate as a power law of the relative energy density of dark energy (taken to evolve as $\Lambda$). We combine the lensing datasets with BAO and Supernovae data which fix the expansion history at late times to constrain the free parameter of our model: the coefficient $\beta$, see Eq.~(\ref{eq:alpha}).  

Our main findings are as follows: 
\begin{itemize}
    \item Our one-parameter DETG model provides an improved fit to galaxy lensing and CMB lensing.  We find $2.5-3\sigma$ evidence for non-zero $\beta$ in our combined fit, which includes BAO and SN data. 
    \item Galaxy lensing favors slower growth at late times (a positive value of $\beta$). Since no reference to an external amplitude calibration is used in the fit, this is evidence for slower growth, independent of the usual $S_8$ tension.
    \item CMB lensing shows a weak preference for {\it stronger} growth at late times. This preference arises from the shape dependence of its power spectrum. Galaxy lensing on the other hand has a preference for slower growth. However, this difference is not significant, and the best-fit model (which has slower growth) improves the overall $\chi^2$ by about 7, with only one additional free parameter. 
    \item Our model does not explicitly allow for scale-dependent modifications to the growth (beyond the one induced in angular statistics by scale projection of a modified time dependence). The success of the time dependence we introduce suggests that current data does not require this additional feature, but to understand the origin of a possible deviation from $\Lambda$CDM, it is worth exploring scale dependence as well, especially as measurements improve. 
\end{itemize}

There are several caveats and extensions to our simple study. 
We have used new publicly available ACT lensing power spectra but not the cross-correlation with galaxy lensing (for which we use the cross-correlations with SPT lensing \cite{DES:2022xxr}). This measurement would have higher signal-to-noise than the measurements used here and could therefore also impact the results. On the galaxy lensing side, we have used the fiducial DES Y3 measurement of lensing two-point correlations. Other studies, using galaxy clustering, Fourier space correlations or higher order statistics, or measurements from KiDS or HSC SSP \cite{DES:2021lsy,DES:2022qpf,DES:2021epj,KiDS:2020suj,Dalal:2023olq}, may yield somewhat different results. 
Rather than carry out an exhaustive comparison, in future work we will use measurements from the full DES survey (the Year 6 data) and other galaxy and CMB surveys to improve our constraints. 
Higher redshift perturbation tracers, such as Lyman-$\alpha$ forest and the future Dark Energy Spectroscopic Instrument (DESI) data, would be useful to further test our model.
Finally, as noted above, our model gives only limited freedom to the time dependence of the growth factor, a more extensive exploration that includes scale-dependent deviations is merited, especially with improved measurements expected from DES, DESI, ACT, and South Pole Telescope in the near future. For a study of scale-dependent modifications due to baryonic physics, see e.g.~\cite{Amon:2022azi,Preston:2023uup}.

The model in this paper is a phenomenological model. It could possibly be realized by some physical models, e.g. ones that involve a significant DE clustering; modifications to gravity at late times \cite{Pogosian:2021mcs, Raveri:2021dbu}; interactions between dark energy and dark matter~\cite{Pourtsidou:2016ico,Poulin:2022sgp}.
We leave the investigation of these interesting model realizations to future work.

\begin{acknowledgments}
\textbf{Acknowledgements:}
We thank Neal Dalal and 
Mathew Madhavacheril
for the helpful discussions and comments.
Computing resources were provided by the National Energy Research Scientific Computing Center (NERSC), a U.S. Department of Energy Office of Science User Facility operated under Contract No. DE-AC02-05CH11231, and by the University of Chicago Research Computing Center through the Kavli Institute for Cosmological Physics. 
M-X.L. is supported by funds provided by the Center for Particle Cosmology. B.J. and M.G. are supported in part by the US Department of Energy grant DE-SC0007901.
M.R. acknowledges financial support from the INFN InDark initiative.
S.L. is supported under a contract with the National Aeronautics and Space Administration and funded through the internal Research and Technology Development program.
J.M. is supported in part by the Government of Canada through the Department of Innovation, Science and Economic Development and by the Province of Ontario through the Ministry of Colleges and Universities. 
\end{acknowledgments}

\bibliography{ref}

\end{document}